\documentclass[CRPHYS,Unicode,manuscript]{cedram}

\usepackage{braket}

\title{Persistent currents in a strongly interacting multicomponent Bose gas on a ring}

\author{\firstname{Giovanni} \middlename{} \lastname{Pecci}\CDRorcid{0000-0002-9081-8361}\IsCorresp}
\address{Université Grenoble-Alpes, CNRS, LPMMC, 38000 Grenoble, France}
\email[G. Pecci]{giovanni.pecci@lpmmc.cnrs.fr}
\CDRGrant[ANR]{21-CE47-0009}
\thanks{We acknowledge support from the Quantum-SOPHA ANR Project N. ANR-21-CE47-0009} 

\author{\firstname{Gianni} \middlename{} \lastname{Aupetit-Diallo}\CDRorcid{0000-xxx-0000-yyyy}\IsCorresp}
\address{Université Côte d’Azur, CNRS, Institut de Physique de Nice, 06560 Valbonne, France}
\email[G. Aupetit-Diallo]{gianni.aupetit-diallo@inphyni.cnrs.fr}

\author{\firstname{Mathias} \middlename{} \lastname{Albert}\CDRorcid{0000-0002-6593-0300}\IsCorresp}
\addressSameAs{2}{}
\email[M. Albert]{mathias.albert@inphyni.cnrs.fr}

\author{\firstname{Patrizia} \middlename{} \lastname{Vignolo}\CDRorcid{0000-xxx-0000-yyyy}\IsCorresp}
\addressSameAs{2}{}
\email[P. Vignolo]{patrizia.vignolo@inphyni.cnrs.fr}

\author{\firstname{Anna} \middlename{} \lastname{Minguzzi}\CDRorcid{0000-0003-3531-8759}\IsCorresp}
\addressSameAs{1}{}
\email[A. Minguzzi]{anna.minguzzi@lpmmc.cnrs.fr}

\ESM{Supplementary material for this article is supplied as a separate 
archive available from  the journal's
website under 
article's URL 
or from the author.}

\keywords{Quantum gases, one-dimensional systems, strong interactions, artificial gauge fields}

\begin{abstract} 
We consider a two-component Bose-Bose mixture at strong repulsive interactions in a tightly confining, one-dimensional ring trap and subjected to an artificial gauge field. By employing the Bethe Ansatz exact solution for the many-body wavefunction, we obtain the ground state energy and the persistent currents. For each value of the applied flux, we then determine the symmetry of the state under particles exchange. We find that the ground-state energy and the persistent currents display a reduced periodicity with respect to the case of non-interacting particles, corresponding to reaching states with fractional angular momentum per particle. We relate this effect to the change of symmetry of the ground state under the effect of the artificial gauge field. Our results generalize the ones previously reported for fermionic mixtures with both attractive and repulsive interactions and highlight the role of symmetry in this effect.
\end{abstract}

\begin{document}

\maketitle

\section{Introduction}
Ultracold atomic gases are a very versatile system for investigating fundamental physics and for quantum simulation \cite{RevModPhys.80.885,doi:10.1126/science.aal3837,lewenstein2012ultracold}. The steady progress in trapping and manipulating cold atoms allows for an unprecedented control on the parameters of the system such as interaction strength, number of particles and components, and geometry \cite{RevModPhys.82.1225,doi:10.1126/science.1201351}. In particular, cold atoms can be confined in one-dimensional potentials of different geometries and subsequently used to experimentally realize \cite{paredes2004tonks,doi:10.1126/science.1100700,kinoshita2006quantum,https://doi.org/10.48550/arxiv.2202.11071} paradigmatic strongly correlated one-dimensional models \cite{PhysRev.130.1605,PhysRevLett.19.1312,gaudin_bethe_2014,sutherland1968further,sutherland1975model}. 
One of the main advantages of investigating these systems rely on the wide class of exact methods one can implement. For instance, one-dimensional homogeneous systems interacting via zero range potential can be solved at any interaction strength by means of Bethe Ansatz \cite{gaudin_bethe_2014,sutherland_beautiful_2004,Oelkers_2006}, while models including non-homogeneous confinement can be exactly solved in some specific interaction regimes using different methods \cite{minguzzi2022strongly,volosniev_strongly_2014,PhysRevA.90.013611}. 

In the context of homogeneous systems, quantum gases trapped in a ring-shaped potential are a suitable platform to investigate quantum coherence and transport \cite{amico2021roadmap}. These systems can be threaded by an artificial gauge flux \cite{dalibard2015introduction,RevModPhys.83.1523}. 
The corresponding artificial gauge field can be implemented e.g.~by stirring the gas using a barrier \cite{wright2013Driving} or by imprinting a geometrical phase to the gas  \cite{Kumar2018producing}. 
The artificial gauge field  induces a persistent current of particles flowing in the ring (see e.g.~ \cite{zvyagin1995persistent,https://doi.org/10.48550/arxiv.2107.08561} for reviews in fermionic and bosonic systems). Persistent currents are a manifestation of quantum coherence of the particles all over the ring. They coincide with supercurrents in the case of superfluid or superconductors, but can also occur in normal fermionic systems, and can be used to probe different phases of the system \cite{PhysRevLett.95.063201}. 
In ultracold atomic rings, the persistent current can be experimentally accessed by co-expansion protocols of the gas on the ring and a reference gas at the center. The value and the sign of the current emerges as the result of spiral interferometry analysis\cite{PhysRevX.4.031052,PhysRevLett.113.135302,PhysRevA.92.033602,PhysRevLett.128.150401,https://doi.org/10.48550/arxiv.2204.06542,https://doi.org/10.48550/arxiv.2206.02807}.  

In analogy with superconducting rings \cite{byers1961theoretical}, the persistent current is a periodic function of the external effective flux, whose period is defined as the quantum of flux of the gas \cite{nanoelectronics1991dk}. The increase the flux by an amount equal to the quantum of flux corresponds to a change of the value of the total angular momentum and consequently of the current. A reduction of the period of the current as a function of the flux  has been predicted both for  Fermi and and for Bose gases with strong attractive interactions\cite{10.21468/SciPostPhys.12.4.138,PhysRevLett.101.106804,PhysRevResearch.3.L032064}, and corresponds to {\it angular momentum fractionalization}, i.e. to the possibility of associating a fractional value of angular momentum per particle.  This phenomenon relies on the formation of two-body and many-body bound states, respectively for Fermi and Bose gases, which deeply affect the state of the gas. In particular, the period of the current oscillations as a function of the effective flux is reduced by a factor corresponding to the number of particles giving rise to the bound state: two for attracting fermions forming pairs and $N$ -- with $N$ being the total number of particles -- for attracting bosons forming the quantum equivalent of a bright soliton. A similar phenomenon has been predicted for a multi-component Fermi gas with very large repulsive interactions \cite{PhysRevB.45.11795,10.21468/SciPostPhys.12.1.033}. In this case,  the period of the persistent current oscillations is also reduced by a factor $N$. However, this phenomenon is not related to the formation of molecules, rather, it is due  to the creation of fermionic spin excitations, i.e spinons, in the ground state at finite flux \cite{PhysRevB.45.11795,10.21468/SciPostPhys.12.1.033}.

In this article, we study a strongly-interacting two-component Bose gas trapped on a ring and threaded by an artificial gauge field. The model is exactly solvable: using the Bethe Ansatz we explicitly obtain the many-body wavefunction,  the ground-state energy, and the persistent current at very large interactions up to four particles. In analogy with fermionic mixtures, we find that at large interactions the period of the ground-state energy and of the persistent current as a function of the flux is reduced by a factor equal to the total number of particles. At non-zero flux, such ground-state branches correspond to spin-excited states at zero flux, characterized by a different value of angular momentum. 
Furthermore, we characterize the symmetry under particle exchange of the ground state at varying values of the effective flux.    We find that each ground-state branch
has a different symmetry i.e.~it is associated to a different Young tableau. We also show that, when the number of low-energy spin excitations exceeds the number of particles, for some values of the flux the ground state may be degenerate  and  correspond to more than one symmetry under particle exchange. For each of such cases we identify the corresponding Young tableaux. 

In the following, after introducing the model and the definitions, we first consider the instructive case of two particles and then we study the more involved case of a mixture of four particles.

\section{Model and definitions}
We consider a two-component Bose-Bose mixure of $N=N_\uparrow+ N_\downarrow$ particles, focusing on the balanced  case  $N_\uparrow= N_\downarrow$.
The bosons interact via a 
delta potential of strength $g$, taking the case where the interaction strength $g$ of the intra-species interactions is the same as the one of the inter-species interactions.  The gas is confined in a one-dimensional ring of radius $R \doteq \frac{L}{2\pi}$, with $L$ being the circumference of the ring. We consider an artificial gauge field, e.g.~induced by setting the system in rotation with frequency $\Omega$ inducing an effective flux $\Phi= 2 \Omega  \pi  R^2$ flowing through the ring.

The Hamiltonian of the system is:
\begin{equation}
\mathcal{H} = \sum_{j=1}^N \frac{1}{2m} \left(p_j - m \Omega R\right)^2  + g \sum_{j<\ell} \delta(x_j - x_\ell) 
\label{hamiltonian}
\end{equation}
where $m$ is the mass of the particles. 
In the following, we define the quantities $c \doteq \frac{2m}{\hbar^2}g$ and $\tilde{\Phi} \doteq \frac{\Phi}{\Phi_0}$, where $\Phi_0 = \frac{h}{m}$, to indicate respectively the interaction strength and the reduced  flux. The kinetic part of the Hamiltonian can be hence rewritten as 
$\mathcal{H}_{kin} = \sum_{j=1}^N \frac{1}{2m} \left (p_j - \frac{2 \pi\hbar }{L} \tilde \Phi\right)^2 $.
We also set $\epsilon = \frac{\hbar^2 \pi^2}{mL^2}$ as the energy scale. 

This model is integrable at any interaction strength and can be solved exactly using Bethe Ansatz \cite{Li_2003,Oelkers_2006,PhysRevA.73.021602}. In each coordinate sector $Q = \{x_{Q(1)} \leq x_{Q(2)} \leq ... x_{Q(N)} \} $ the wavefunction reads \cite{Oelkers_2006}:
\begin{equation}
\Psi(x_{Q(1)}..x_{Q(N)}) = \sum_P A_Q(\Lambda_m, k_{P(j)},c) \exp \Bigl\{i \sum_P k_{P(j)} x_{Q(j)} \Bigr\} ,
\label{wavefunction}
\end{equation}
where the sum is performed over all the possible permutations $P$ in the symmetric group $S_N$. In Eq.(\ref{wavefunction}) we introduced the amplitudes $A_Q$, the \textit{spin rapidities} $\Lambda_m$ with  $m=1..M$ ,  $M$ being the number of spin down particles, and the \textit{charge rapidities} $k_j$ with  $j=1..N$. The two sets of rapidities fully specify  the wavefunction of the system: they can be obtained for each value of $\tilde{\Phi}$ by solving the coupled Bethe equations \cite{Li_2003},
\begin{equation}
\begin{cases}
L k_j = 2\pi \mathcal{I}_j + 2 \sum_{\ell=1}^N \arctan\Biggl(\frac{k_\ell - k_j}{c}\Biggr) + 2 \sum_{m=1}^M \arctan\Biggl(\frac{2(k_j - \Lambda_m)}{c}\Biggr) \\
\sum_{j=1}^N 2\arctan \Biggl(\frac{2}{c} (\Lambda_m - k_j) \Biggr) = 2\pi \mathcal{J}_m + \sum_{n=1}^M 2 \arctan \Biggl(\frac{\Lambda_m - \Lambda_n}{c}\Biggr),
\label{bethe_eqs}
\end{cases}
\end{equation}
where we introduced the charge and the spin quantum numbers $\mathcal{I}_j$ and $\mathcal{J}_m$, which are integers or half-integers respectively if $N-M$ is odd or even. The energy of the system is given by $E(\tilde{\Phi}) = \frac{\hbar^2}{2m}\sum_j (k_j - \frac{2\pi}{L}\tilde{\Phi})^2$ and the total momentum is  $P = \hbar \sum_j k_j$.

The choice of the quantum numbers fixes the state of the system. In particular, in the ground state, adjacent quantum numbers are spaced by one unit. They are chosen, for each value of the flux, such that the corresponding rapidities $k_j$ minimize the energy $E(\tilde{\Phi})$ \cite{Li_2003,IMAMBEKOV20062390,essler_2005}. 

In this article, we focus on the Tonks-Girardeau (TG) fermionized limit $c \to \infty$, where the inter-particle interactions are infinitely repulsive. This induces an effective Pauli principles among the particles: the wavefunction of the system vanishes as two particles occupy the same spatial position, still ensuring the preservation of  the bosonic symmetry under particle exchange. In the TG regime, the Bethe Ansatz solution of the model is markedly simplified. To describe such limit,  we introduce the rescaled spin rapidities $\lambda_m \doteq \frac{2\Lambda_m}{c}$ \cite{Oelkers_2006,PhysRevB.41.2326,essler_2005}, which we assume to be finite in the limit $c\to \infty$. In this limit, exploiting the anti-symmetry of the arctangent function, 
the Bethe equations read:
\begin{equation}
\begin{cases}
L k_j = 2\pi \Bigl( \mathcal{I}_j - \frac{1}{N} \sum_{m=1}^M \mathcal{J}_m\Bigr)\\
2N \arctan (\lambda_m) = 2\pi  \mathcal{J}_m + \sum_{n=1}^M 2 \arctan (\lambda_m - \lambda_n).
\label{bethe_eqs_infinite}
\end{cases}
\end{equation}
The first equation fixes the energy of the system: in this interaction regime the distribution of the quantum numbers $\mathcal{J}_m$ - thus the spin excitations - affects the total momentum and the kinetic energy.  
The second equation coincides with the Bethe equations for an isotropic spin chain \cite{essler_2005,bethe1931theorie,franchini2017introduction} and does not depend on the charge degree of freedom. The same spin-charge decoupling occurs in the wavefunction (\ref{wavefunction}), where the amplitudes satisfy $\lim_{c\to \infty} A_Q(\Lambda_m, k_{P(j)},c) = A_Q(\Lambda_m/c) = A_Q(\lambda_m) $ and explicitly read \cite{essler_2005, Oelkers_2006}:
\begin{equation}
A_{Q}(\lambda_1,...\lambda_M) \propto (-1)^{\lvert Q\rvert}\sum_R \prod_{1 \leq m < n \leq M} \frac{\lambda_{R(m)} - \lambda_{R(n)} - 2i}{\lambda_{R(m)} - \lambda_{R(n)}} \prod_{l=1}^{M} \Biggl( \frac{\lambda_{R(l)}- i}{\lambda_{R(l)} +i}  \Biggr)^{y_{Q(l)}}.
\label{bethe_amplitudes_infinite}
\end{equation}
In this equation, the integer $y_{Q(l)}$ labels the position of the $l$-th spin down in the coordinate sector $Q$. The notation $\lvert Q \rvert$ indicates the sign of the permutation linking the coordinate sector $Q$ with the identical coordinate sector defined by $x_1 \leq x_2 \leq \dots \leq x_N$.

We stress that, up to a normalization constant, Eq.~ (\ref{bethe_amplitudes_infinite}) has the same functional structure of the Bethe wavefunction of the isotropic Heisenberg spin chain \cite{bethe1931theorie,franchini2017introduction,essler_2005}.  Despite the same Bethe equations and a similar structure of the spin component of the wavefunction, the expression for the spectra of Hamiltonian (\ref{hamiltonian}) and of the spin chain are in general different:  for the full model (\ref{hamiltonian}) the energy is given by the charge rapidities, while for the spin chain it is linked to the spin rapidities. Still, the correction to first order in $1/c$ of the spectrum of the full model  can be mapped onto the spectrum of the Heisenberg chain by a suitable definition of an effective coupling $J$ of the chain \cite{PhysRevB.45.11795,10.21468/SciPostPhys.12.1.033}.


In order to obtain explicit values for the amplitudes $A_Q$, we solve the Bethe equations (\ref{bethe_eqs_infinite}) \cite{essler_2005,franchini2017introduction} for all the possible distributions of the quantum numbers $\mathcal{I}_j$ and $\mathcal{J}_m$, which are in turn fixed by the number of particles $N$ and the number of down spins $M$. Thanks to the periodicity of the Bethe equations, the number of possible sets of quantum numbers yielding independent solutions of Eq.~(\ref{bethe_eqs_infinite}) is finite. 
In particular, in order to determine the amplitudes of the ground state wavefunction for a fixed value of the flux, we solve the Bethe equation by choosing the sets of quantum numbers that minimize the energy.

The determination of the amplitudes $A_Q$ for each possible spin ordering $Q$ allows us to write explicitly the many-body wavefunction $\Psi(x_{Q(1)}..x_{Q(N)})$ in each coordinate sector and to characterize the symmetry of the state under particle exchange.  The symmetry of the wavefunction is evaluated by computing the expectation value of the $S_N$ class-sum operator $\Gamma_2 \doteq \sum_{a<b} P_{ab}$ on the state itself \cite{Decamp_2016, IMAMBEKOV20062390,RevModPhys.55.331}, $P_{ab}$ being the permutation operator acting on all the $N$ particles. 
The knowledge of the class-sum operator allows us to determine the Young tableaux associated to the state. In particular, the eigenvalues $\gamma_2$ of the $\Gamma_2$ operator are linked to the Young tableaux encoding the possible symmetries of the state through the following relation:
\begin{equation}
    \gamma_2 = \frac{1}{2} \sum_j n_j (n_j - 2j + 1),
\end{equation}
where $j$ labels the line of the corresponding Young tableau and $n_j$ is the number of boxes in the $j$-th line. 
In the following, we use the convention that the horizontal Young tableau corresponds to the fully symmetric state, while the vertical tableau corresponds to the most anti-symmetric state.

\section{Case of two particles}
Before tackling the case of larger numbers of particle, it is instructive to understand the solution for $N=2$ particles, i.e one boson for each component of the mixture. When $N=2$ and $M=1$ there are only two coordinate sectors, namely $\mathbb{1}: x_1 \leq x_2 $ and $\tilde {Q}: x_2 \leq x_1$. Without loss of generality we assume the positions of the spin down particle in the two sectors to be $y_{\mathbb{1}} = 2$ and $y_{\tilde Q} = 1$. We use Eq.~(\ref{bethe_amplitudes_infinite}) to compute the amplitudes of the Bethe wavefunction:
\begin{align}
&A_{\mathbb{1}}(\lambda) \propto \Biggl( \frac{\lambda - i}{\lambda +i}  \Biggr)^2, \notag \\
&A_{ \tilde Q}(\lambda) \propto (-1) \Biggl( \frac{\lambda - i}{\lambda +i}  \Biggr),
\label{bethe_amplitudes_N2}
\end{align}
where we used the property $\lvert \mathbb{1} \rvert = 0$ and $\lvert \tilde Q \rvert = 1$. The Bethe equation for the spin rapidity $\lambda$ reads:
\begin{equation}
\arctan(\lambda) = \frac{\pi}{2} \mathcal J, \quad \mathcal J \in \mathbb{Z}.
\end{equation}
In the ground state at zero flux we have  $\mathcal J=0$ and $A_{\mathbb{1}}(0) = A_{ \tilde Q}(0) = 1$, while in the first excited state $\mathcal J=1$ and consequently $A_{\mathbb{1}}(\infty) = 1, A_{ \tilde Q}(\infty) = -1$. Next, in order to make a link with the standard solutions in the fermionized limit, we decompose the wavefunction in each coordinate sector on a basis of anti-symmetric combinations of plane waves, as opposed to Eq.~(\ref{wavefunction}). Consequently, we introduce the amplitudes $\mathcal{A} = (-1)^{\lvert Q \rvert} A_Q$. Explicitly, the wavefunction in this form reads:
\begin{equation}
\Psi(x_1,x_2)= {\mathcal A}(x_1-x_2) \det \left(e^{i k_j x_\ell}\right).
    \label{eq:N2-general}
\end{equation}
A simple reorganization of the terms entering in the above equation yields 
\begin{equation}
    \Psi(x_1,x_2)= {\mathcal A}(x_1-x_2)\sin{\left(  k (x_1-x_2) \right)}e^{iK \left(x_1+x_2\right)/2}
    , \label{psi 2p}
\end{equation}
where $K=k_1+k_2$ is the center of mass momentum and $k=(k_1-k_2)/2$ is the relative momentum of the two-particle system. In the homogeneous ring, there is complete factorization between  the internal structure of the state, encoded in the term  $ {\mathcal A}(x_1-x_2)\sin{k((x_1-x_2))}$, where  ${\mathcal A}(x_1-x_2)$ controls  the overall symmetry under exchange of particles, 
and the center-of-mass part $\exp(iK (x_1+x_2)/2)$. The latter is the one which couples to the artificial gauge flux \cite{manninen_quantum_2012,10.21468/SciPostPhys.12.4.138}.
The corresponding value of the  energy is  $E=\sum_{j=1,2} (\hbar^2/2m) [k_j-(2 \pi/L)\tilde \Phi]^2$.

The values of the wave-vectors $k_1$, $k_2$ are obtained by imposing the periodic boundary conditions
$\Psi(x_1+L,x_2)=\Psi(x_1,x_2)=\Psi(x_1,x_2+L)$. 
The function $ \mathcal A(x_1-x_2)$ for the ground state depends on the value of the artificial gauge field.

For $-0.25<\tilde \Phi<0.25$ the ground state of distinguishable bosons coincides with the one of identical TG bosons i.e. $\mathcal{A}(x_1-x_2)={\rm sign} (x_1-x_2)$. In this case the periodic boundary conditions imposed on Eq.(\ref{eq:N2-general}) yield $k_1+k_2=(2\pi/L) 2 p $ and $k_1+k_2=(2\pi/L) q$ with $p$, $q$ integers. The solution for the ground state gives $k_1=-\pi/L$ and $k_2=\pi/L$. 

Notice that thanks to the analogy of the Hamiltonian with the one of particles in a crystal with quasi-momentum $\tilde \Phi$, the same choice for $\mathcal{A}(x_1-x_2)$ holds for all intervals of flux obtained by a translations of the interval $-0.25<\tilde \Phi<0.25$  by integer numbers, i.e shifting $\Phi$  by integer multiples of $\Phi_0$.
 
For $0.25<\tilde \Phi<0.75$ the ground state is instead obtained by choosing  $\mathcal{A}(x_1-x_2)=1$, as for spinless fermions. This corresponds to the Bethe Ansatz solution for the wavefunction of the first excited state at zero flux. In this case, the periodic boundary conditions yield $k_1=0$ and $k_2=2\pi/L$. As above, the same choice for $\mathcal{A}(x_1,x_2)$ holds for all intervals of flux values obtained by translations of the considered interval  by integer numbers.

\begin{figure}
    \centering
    \includegraphics[scale=0.4]{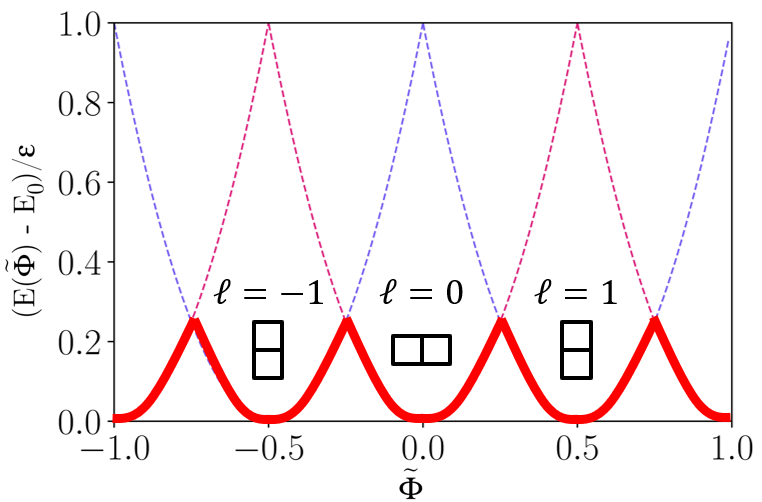}
    \caption{Ground state energy (relative to the energy $E_0$ at zero flux, in units of $\epsilon$) as a function of the reduced flux (dimensionless) for the case $N=2$, $M=1$ (red solid line). The magenta and violet dashed lines  correspond to the energy landscape for $N=2$ single-component Tonks-Girardeau bosons and spin-polarized fermions respectively.  The total angular momentum quantum number and the Young tableau indicating the symmetry of the ground state under particle exchange are also indicated on each ground-state branch.}
    \label{fig1p1}
\end{figure}

By collecting all the above considerations, we obtain the ground-state energy as a function of flux (see Fig.~\ref{fig1p1}): it consists of piece-wise parabolas, with half periodicity with respect of the  flux quantum $\Phi_0$.  We notice that each parabola is associated to a different value of the total momentum $P=\hbar(k_1+k_2)$, and hence of the total angular momentum  $L_z$ along the direction perpendicular to the ring plane,
labelled by $\ell = \braket{L_z}/\hbar$,  as also indicated on the figure. We notice that the halved periodicity implies \emph{fractional angular momentum per particle} as already reported for the case of attracting bosons \cite{10.21468/SciPostPhys.12.4.138}, paired fermions \cite{PhysRevB.45.11795,PhysRevA.7.2187,PhysRevResearch.3.L032064}, and SU(N) fermionic mixtures \cite{10.21468/SciPostPhys.12.1.033}.

Our explicit solution allows also to readily obtain the symmetry of the ground state. For the parabola centered at zero flux (and all its translations by $\Phi_0$), the wave-function is fully symmetric, while for the one centered at $\Phi_0/2$ (and all its translations by $\Phi_0$) the wave-function is fully anti-symmetric. The corresponding Young Tableaux are also depicted in Fig.~\ref{fig1p1}.

Let us summarize the four main aspects emerging from the analysis of the two-particle case: (i) the ground state of the mixture  on a ring is not degenerate, at difference from the case of a mixture under harmonic confinement \cite{volosniev_strongly_2014}, (ii) the ground-state energy as a function of the flux is given by piece-wise parabolas, each of them characterized by a given value of total angular momentum specified by $\ell$, (iii) each parabola has a well-defined symmetry (either fully symmetric or fully anti-symmetric), and (iv) the case of a two-component mixture displays a halving of the periodicity with respect to the case of a spin-polarized Fermi gas (parabolas centered at semi-integer values of $\Phi_0$) as well as the one of a single-component TG gas (parabolas centered at integer multiples of $\Phi_0$).

 In the following, we will treat the more challenging case of  a 2+2 spin mixture.


\section{Results for $N = 4, \quad M = 2$}
In this section, we provide the results for a balanced multicomponent Bose gas of $N = 4$ particles and $ M = 2$ spins down. The quantum numbers $\mathcal{I}_j$ and $\mathcal{J}_m$ are both semi-integers \cite{Li_2003}. We can write Eq.~(\ref{bethe_amplitudes_infinite}) as follows:

\begin{equation}
A_Q (\lambda_1, \lambda_2)\propto (-1)^{\lvert Q\rvert} \Biggl(\frac{\lambda_1 - \lambda_2 - 2i}{\lambda_1 - \lambda_2} \Bigl( \frac{\lambda_1- i}{\lambda_1 +i} \Bigr)^{y_{Q(1)}}\Bigl( \frac{\lambda_2- i}{\lambda_2 +i} \Bigr)^{y_{Q(2)}} + \frac{\lambda_2 - \lambda_1 - 2i}{\lambda_2 - \lambda_1} \Bigl( \frac{\lambda_2- i}{\lambda_2 +i} \Bigr)^{y_{Q(1)}}\Bigl( \frac{\lambda_1- i}{\lambda_1 +i} \Bigr)^{y_{Q(2)}}\Biggr).
\label{amplitudes2+2}
\end{equation}

\begin{table}[ht]
\caption{Solutions of the Bethe equations for a strongly repulsive Bose-Bose mixture of $N=4$ particles and $M=2$ spin-down particles. We consider various values of total angular momentum $P$ and quantum numbers configuration $\mathcal J_1, \mathcal J_2$. Such solutions are the ground-state and the first excited states for zero reduced flux $\tilde \Phi$ (see column $E(0)/\epsilon$ ), but become the ground state in a given interval of flux as indicated in the last column of the table.} 
\centering 
\begin{tabular}{c c c c c c} 
\hline \hline  
$\rule{0pt}{2.5ex}    
 P$  &$\mathcal{J}_1+\mathcal{J}_2$ &$\tilde{\lambda}_1$  & $\tilde{\lambda}_2$ &$E(0)/\epsilon$ & Reduced flux interval \\ [0.5ex]
\hline 
$0$       &0  & $1/\sqrt{3}$  & $-1/\sqrt{3}$  &$10$ &$-1/8 \leq \tilde \Phi \leq 1/8$ \\
$0$       &0  & $i\infty$  & $-i\infty$  &$10$&$-1/8 \leq \tilde \Phi \leq 1/8$  \\ 
$2\pi/L$  &1  & $-1$  & $\infty$  &$21/2$ &$1/8 \leq \tilde \Phi \leq 3/8$ \\ 
$4\pi/L$  &2  & $0$  & $\infty$  &$12 $ &$3/8 \leq \tilde \Phi \leq 5/8$ \\ 
$4\pi/L$  &2  & $i$ & $-i$  & $12 $& $3/8 \leq \tilde \Phi \leq 5/8$ \\ 
$6\pi/L$  &3  & $1$  & $\infty$  &$29/2$ &$5/8 \leq \tilde \Phi \leq 7/8$ \\ 
[1ex] 
\hline
\end{tabular}
\label{table:solutionN4M2}
\end{table}

The set of Bethe equations is:
\begin{equation}
\begin{cases}
Lk_j = 2 \pi \mathcal{I}_j - \frac{2\pi}{4} (\mathcal{J}_1+\mathcal{J}_2) \\
8 \arctan(\lambda_1) = 2\pi \mathcal{J}_1 - \arctan \frac{\lambda_2 - \lambda_1}{2} \\
8 \arctan(\lambda_2) = 2\pi \mathcal{J}_2 + \arctan \frac{\lambda_2 - \lambda_1}{2} ,
\end{cases}
\end{equation}
which can be simplified using the trigonometric relation $\arctan(a) + \arctan(b)= \arctan(\frac{a+b}{1-ab})$ 
\begin{equation}
\begin{cases}
Lk_j = 2 \pi \mathcal{I}_j - \frac{2\pi}{4} (\mathcal{J}_1+\mathcal{J}_2)  \\
\frac{\lambda_1+\lambda_2}{1-\lambda_1 \lambda_2} = \tan\bigl(\frac{\pi}{4}(\mathcal{J}_1 + \mathcal{J}_2) \bigr)\\
8 \arctan(\lambda_2) = 2\pi \mathcal{J}_2 + \arctan \frac{\lambda_2 - \lambda_1}{2} .
\end{cases}
\label{bethe_eqs_N4}
\end{equation}
In order to minimize the energy associated to the charge sector we have to minimize  $\sum_j\mathcal{I}_j$. As a consequence, the set of quantum numbers $\mathcal{I}_j$ for the ground state of the charge sector is $\mathcal{I}_j = \{-\frac{3}{2}, -\frac{1}{2},\frac{1}{2},\frac{3}{2}\}$. 
Moreover, due to the periodicity of the tangent function, the second equation only gives independent solutions for $(\mathcal{J}_1 + \mathcal{J}_2) (\text{mod} \ 4)$. Therefore, we can focus on the four cases $\mathcal{J}_1 + \mathcal{J}_2 = 0,1,2,3$, which, for each value of $\tilde{\Phi}$, correspond to respectively the ground state 
and the first three excited states. Explicitly, these configurations yield the following values for the total momentum 
$P= 0, \frac{2\pi}{L}, \frac{4\pi}{L}, \frac{6\pi}{L}$. The solutions of Eq.~(\ref{bethe_eqs_N4}) are listed in Table \ref{table:solutionN4M2}. Remarkably, if we allow for complex $\lambda_n$, multiple solutions can be associated to the same value of the momentum. We define $a_Q^{\ell,i} =  A_Q (\tilde \lambda_1^i(P), \tilde \lambda_2^i(P))$ as the amplitudes of the Bethe wavefunction for each configuration of quantum numbers and where $\tilde{\lambda}_{1,2}^i(P)$ are the $i$-th solutions of the last two Bethe equations (\ref{bethe_eqs_N4}) for a fixed value of the total momentum $P$, labelled by the quantum number $\ell$. In particular, for $P=0$ and $P= 4\pi/L$ we get two solutions, while for $P=2\pi/L$ and $P=6\pi/L$ the solution is unique. We also stress that in order to get all the possible low-energy excitations, we had to include singular solutions of the Bethe equations \cite{Nepomechie_2013,Kirillov_2014}.

\begin{table}[ht]
\caption{Amplitudes $a_Q^{\ell,i}$ corresponding to the different spin sectors for $N=4$ and $M=2$.}
\centering 
\begin{tabular}{c c c c c c c } 
\hline \hline
\rule{0pt}{2.5ex} Sector  &$a_Q^{(0,1)}$ & $a_Q^{(0,2)}$ & $a_Q^{(1,1)}$  & $a_Q^{(2,1)}$ & $a_Q^{(2,2)}$  & $a_Q^{(3,1)}$ \\ [0.5ex]
\hline 
$\ket{\uparrow \uparrow \downarrow \downarrow}$  & $2$  & $2$  &$1-i$ &$0$ &$2$ &$1+i$	\\
$\ket{\uparrow \downarrow \uparrow \downarrow}$  & $-4$ & $2$  &$0$  &$2$		&$0$ &$0$\\ 
$\ket{\downarrow \uparrow \uparrow \downarrow}$  & $2$  & $2$  &$1+i$  &$0$	&$-2$ &$1-i$\\
$\ket{\uparrow \downarrow \downarrow \uparrow}$  & $2$  & $2$  &$-1-i$ &$0$	&$-2$ &$-1+i$\\
$\ket{\downarrow \uparrow \downarrow \uparrow}$  & $-4$ & $2$  &$0$ &$-2$		&$0$ &$0$\\ 
$\ket{\downarrow \downarrow \uparrow \uparrow}$  & $2$  & $2$  &$-1+i$ &$0$	&$2$ &$-1-i$\\  [1ex] 
\hline
\end{tabular}
\label{table:apsN4M2}
\end{table}

In Table \ref{table:apsN4M2} we show all the possible $a_Q^{\ell,i}$ for this case in the different coordinate sectors, defined by the possible spin orderings. We get six possible solutions, which correspond to six different states. This value coincides with the possible and distinguishable spin configurations allowed in this case, given in general by $\frac{N!}{M!(N-M)!}$.

\begin{figure}[tbp]
\includegraphics[width=0.8\textwidth]{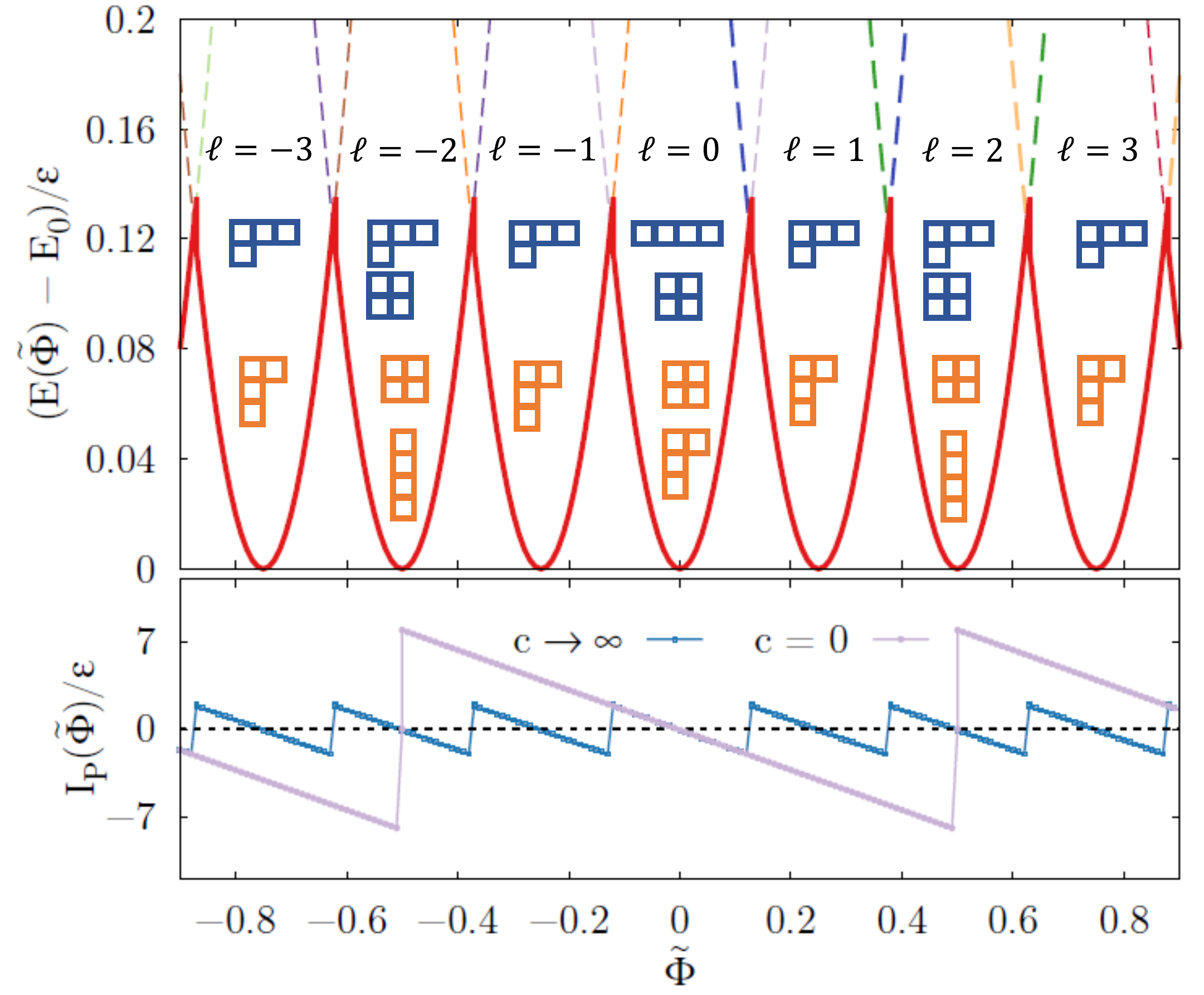}
\caption{Top panel: energy levels (relative to the energy $E_0$ at zero flux, in units of $\epsilon$) as a function of the reduced flux (dimensionelss)  for both a bosonic and a fermionic  mixture with $N=4$ and $M=2$. The red continuous line highlights the ground state of the system. For each value of the flux, we indicate the angular-momentum quantum number of the ground state. The two upper lines of Young tableaux (blue diagrams) indicate the symmetry of the bosonic ground state as function of the flux. The two bottom lines (orange diagrams) the ones of the fermionic ground state. Bottom panel: corresponding persistent current, displaying the $1/N$-periodicity emerging at strong interactions. }
\label{fig:parabolas}
\end{figure}

In the top panel of Fig.~\ref{fig:parabolas} we show the energy $E(\tilde{\Phi})$ as a function of the flux. The continuous red line highlights the ground state energy, which is a periodic function of the gauge flux. Remarkably, the period is reduced by a factor of $N=4$ if compared to the one of the non-interacting case. This effect is also reflected in the persistent current $I_P(\tilde{\Phi}) = - \frac{\partial E}{\partial \tilde{\Phi}}(\tilde{\Phi})$ evaluated in the ground state, which is shown in the bottom panel of Fig.\ref{fig:parabolas}. We compare the persistent current for $c = 0$ and $c \to \infty$. We see the emergence of the $1/N$-reduction of the periodicity. This effect was also reported in strongly interacting Fermi mixtures for repulsive interactions \cite{PhysRevB.45.11795, 10.21468/SciPostPhys.12.1.033}. 

Looking at Fig.\ref{fig:parabolas}, we see that each time the flux increases by $\tilde{\Phi}/N$, the ground state carries a different value of total momentum $P$, corresponding to an angular momentum of $\braket{L_z} = PR$. 

We evaluated the symmetry of the states listed in Table~\ref{table:apsN4M2} by computing the expectation value $\bra{a^{\ell,i}}\Gamma_2 \ket{a^{\ell,i}}$, $\ket{a^{\ell,i}}$ being the vector collecting the coefficients $a_Q^{\ell,i}$ in the different coordinate sectors for the $i$-th state of total momentum $P$ (i.e the columns of Table \ref{table:apsN4M2}), suitably normalized. For each of the above states, this expectation value coincides with an eigenvalue of the class-sum operator $\Gamma_2$, i.e each state has well-defined symmetry. This allows us to link them to a Young tableau and therefore, for any value of the reduced flux, to determine the symmetry of the ground state. In the top panel of Fig.\ref{fig:parabolas} the upper line (blue) of  tableaux provides the symmetry of the ground state for each branch of the ground-state energy as a function of the flux. 

It is instructive to compare our results for the Bose-Bose mixture with the ones for a  Fermi-Fermi mixture with repulsive contact inter-component interactions. In this case, the wavefunction has still the form Eq.~(\ref{wavefunction}). However, the Bethe equations are different since there is no contact interactions among fermions belonging to the same component, and also the symmetry under exchange of particles belonging to the same component is different.  At strong repulsive  interactions the first of Bethe equations (\ref{bethe_eqs_infinite}) reads
$L k_j = 2\pi \Bigl( \mathcal{I}_j + \frac{1}{N} \sum_{m=1}^M \mathcal{J}_m\Bigr)$
while the equation for the spin rapidities coincide to the one for the bosonic case  \cite{PhysRevB.45.11795,10.21468/SciPostPhys.12.1.033}. 
The results for the solutions of the Bethe equations for the fermionic case are summarized in Table~\ref{table:aps3partricles}.
In this case the quantum numbers $\mathcal{I}_j$ for $N=4$ and $M=2$ are integers \cite{Oelkers_2006}. In the ground state, we have $\mathcal{I}_j = \{-2,-1,0,1\}$ which implies $\sum_j \mathcal I_j = -2$. The total momentum 
is $P_F = \frac{2\pi \hbar }{L} \bigl( \sum_{j} \mathcal{I}_j + \sum_a \mathcal{J}_a \bigr) \doteq \frac{\hbar}{R}\ell_F$ 
\cite{PhysRevB.41.2326,essler_2005,10.21468/SciPostPhys.12.1.033}.
The energy levels as a function of the flux are the same as for the Bose-Bose mixture.  Similarly,  for a given value of $\tilde{\Phi}$, the angular momentum of the ground state is the same for bosons and fermions.

On the other hand, the symmetry of the ground state is markedly different in the two cases. We evaluate the symmetry of the fermionic ground-state wavefunction by following the same procedure used for the bosonic system. 
 Since the Bethe equation for the spin rapidities is the same as in the bosonic case, the fermionic amplitudes satisfy $a_{Q}^{\ell_F,j} = a_{Q}^{(\ell-2)  (mod 4),j}$, where $\ell_F$ labels fermionic states with different angular momentum. As a consequence, the same value of the total momentum $P_F$ is associated to different spin rapidities in the two cases and therefore to different amplitudes $A_Q$.  As the amplitudes affect the symmetry of the wave-function, the corresponding Young tableaux  are different in the fermionic and in the bosonic case.
 The Young tableaux indicating the symmetries of the fermionic ground state as a function of the flux are displayed in orange in the top panel of Fig.\ref{fig:parabolas}.

To conclude this part, we remark that -- both in the case of bosonic and fermionic mixtures -- different parabolas display different symmetries, reflecting the fact that they correspond to different excited states at zero flux.  

\begin{table}[ht]
\caption{Solutions of the Bethe equations for a strongly repulsive Fermi-Fermi mixture for $N=4$, $M=2$ and for various values of total momentum $P_F$. We also provide  the energy associated to each state at zero flux and the reduced flux interval where each state becomes the ground state.} 
\centering 
\begin{tabular}{c c c c c c} 
\hline \hline
$\rule{0pt}{2.5ex}    
 P_F$  &$\mathcal{J}_1+\mathcal{J}_2$ &$\tilde{\lambda}_1$  & $\tilde{\lambda}_2$ &$E(0)/\epsilon$ & Reduced flux interval \\ [0.5ex]
\hline 
$0$       &2  & $0$  & $\infty$  &$10$ &$-1/8 \leq \tilde \Phi \leq 1/8$ \\ 
$0$       &2  & $i$ & $-i$ &$10$  &$-1/8 \leq \tilde \Phi \leq 1/8$ \\
$2\pi/L$  &3  & $1$  & $\infty$  &$21/2$  &$ 1/8 \leq \tilde \Phi \leq 3/8$ \\
$4\pi/L$  &4  & $\frac{1}{\sqrt{3}}$  & $-\frac{1}{\sqrt{3}}$  & $12 $ &$3/8 \leq \tilde \Phi \leq 5/8$ \\ 
$4\pi/L$  &4  & $\infty$  & $\infty$  &$12 $ &$3/8 \leq \tilde \Phi \leq 5/8$ \\
$6\pi/L$  &5  & $-1$  & $\infty$  &$29/2$ &$5/8 \leq \tilde \Phi \leq 7/8$ \\
[1ex] 
\hline
\end{tabular}
\label{table:aps3partricles}
\end{table}


\section{Summary and conclusions}

We have studied the ground-state properties of a strongly interacting Bose-Bose mixture subjected to an artificial gauge field on a ring.
We have found that  the ground-state energy is a periodic function of the flux, made by piece-wise parabolas. As compared to the case of non interacting particles, the period of the ground-state energy, as well as the one  of the persistent current, is reduced by a factor $N$, with $N$  the total number of particles in the mixture.   We understand the reduction of periodicity as being due to spin excitations, according to the following mechanism: in the absence of artificial gauge field, the spin excitations lie above the ground state. However, the application of a gauge field decreases the values of the energy of such spin-excited states, making them become the ground state in some intervals of reduced flux.

Each parabola of the ground-state energy landscape is associated to a value of the total angular momentum which increases by one quantum by moving from one parabola to the next. Hence, the emergence of such new branches corresponds to states with fractional  angular momentum per particle. This phenomenon was previously reported for the case of  Fermi mixtures with strong repulsive interactions - our analysis proposes yet another system where this same effect  occurs.

Furthermore, we have characterized the symmetry under exchange of particles of such ground-state branches as a function of the flux, and shown that a single Young tableau can be associated to each branch when it is non-degenerate, while more then one tableau is found when the ground state is degenerate. This analysis confirms the role of spin excitations as being responsible of the reduction of periodicity and the emergence of the new parabolic branches which are absent in the non-interacting regime.

Our study contributes to the deep understanding of the spectrum structure and opens the possibility of designing experiments in which particular symmetries, i.e particular spin states, can be selected.

\section*{Acknowledgements}
A.M. and G.P. acknowledge fruitful discussions with L. Amico.  We acknowledge funding from the ANR-21-CE47-0009 Quantum-SOPHA project.

\bibliographystyle{crunsrt}

\nocite{*}

\bibliography{biblio}

\end{document}